\documentclass[a4paper,11pt]{article}
\pdfoutput=1 
\usepackage{jheppub} 
\usepackage[T1]{fontenc} 
\RequirePackage[displaymath]{lineno} 

\usepackage{epsfig}
\usepackage{graphicx}
\usepackage{dcolumn}
\usepackage{bm}
\usepackage{ltablex,booktabs}
\usepackage{overpic}
\usepackage{subfigure}
\usepackage{float}
\usepackage{color}
\usepackage{amsmath}
\usepackage{mathcomp}
\usepackage{mathrsfs}
\usepackage{multirow}
\usepackage{rotating}
\usepackage{amssymb}
\usepackage{gensymb}
\usepackage{amsmath}
\usepackage{tabularx}
\usepackage{epstopdf}
\usepackage{hang}

\newcommand{\BESIIIorcid}[1]{\href{https://orcid.org/#1}{\hspace*{0.1em}\raisebox{-0.45ex}{\includegraphics[width=1em]{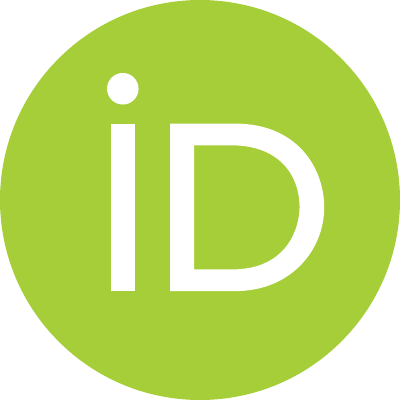}}}}

\title{Search for the radiative decay $D^+_s \to \gamma K^*(892)^+$}
\collaborationImg{\includegraphics[width=0.15\textwidth, angle=90]{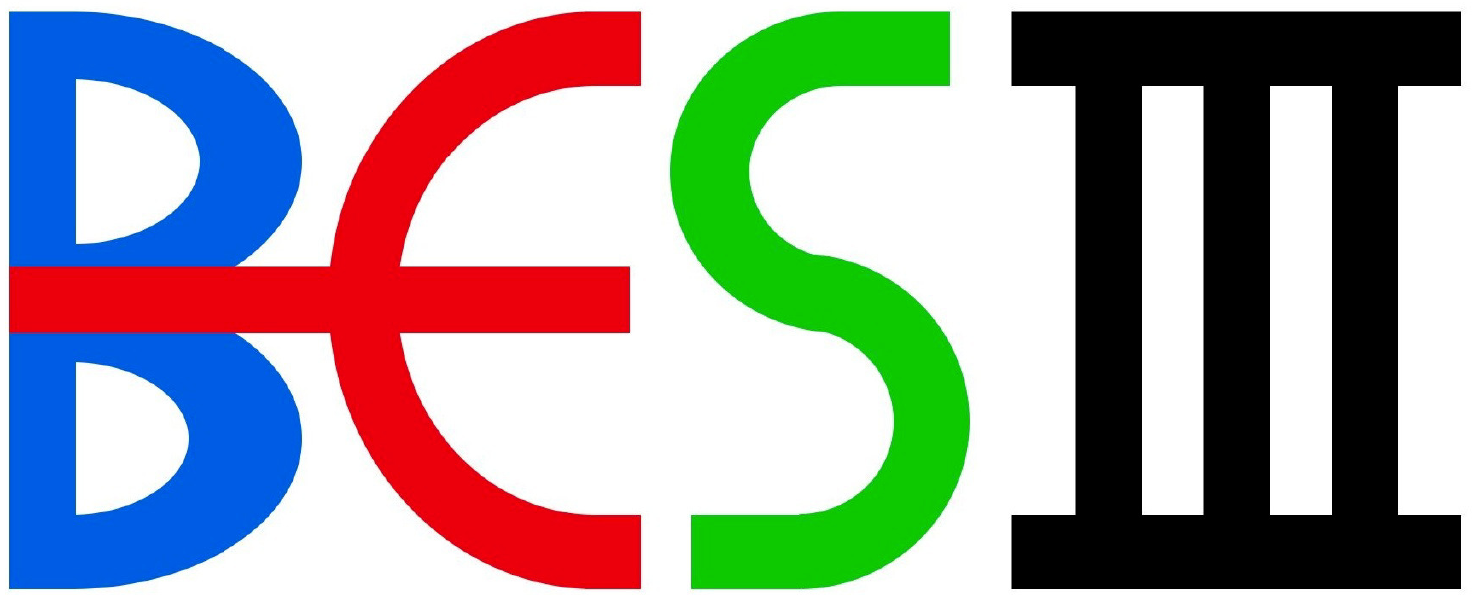}}
\collaboration{The BESIII Collaboration}
\date{\today}
\abstract{Using 7.33 fb$^{-1}$ of $e^+e^-$ collision data samples collected with the BESIII detector at center-of-mass energies between 4.128 and 4.226 GeV,  
we perform a  search for the radiative decay $D_s^+\to\gamma K^*(892)^+$ via $K^*(892)^+\to K^+\pi^0$ and $K^*(892)^+\to K_S^0\pi^+$ for the first time.
No significant signals are observed. The upper limit on the branching fraction of $D^+_s\to\gamma K^*(892)^+$ is set to be $2.3\times10^{-4}$ at the $90\%$ confidence level.}

\arxivnumber{}
\begin{document}
\maketitle
\flushbottom


\section{Introduction}
\begin{figure}[tp!]
\centering	
\includegraphics[width=9.cm]{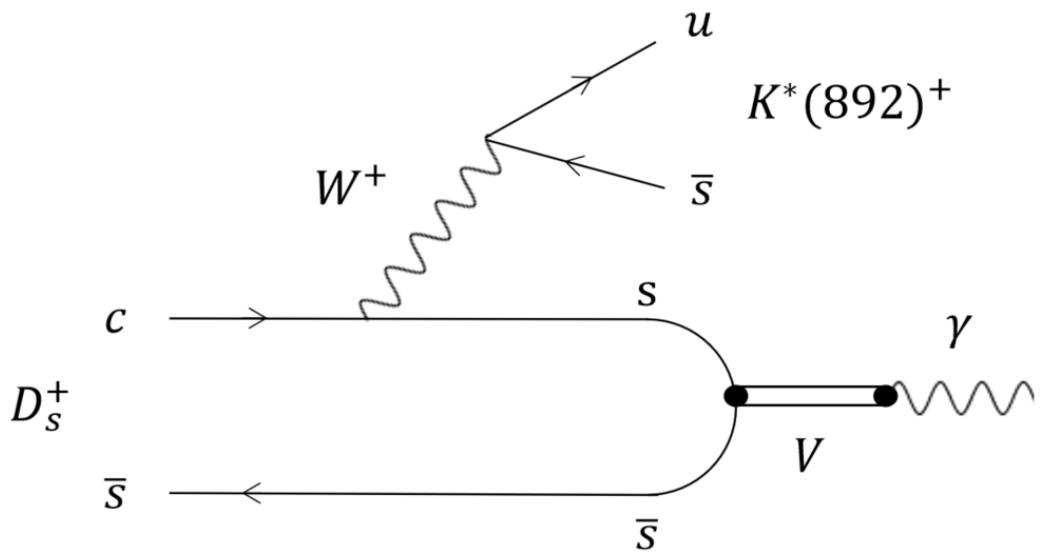}
\caption{The quark diagram of $D^+_s\to \gamma K^*(892)^{+}$.}
\label{fig:feynman1}
\end{figure}

In the weak radiative decays of charmed mesons, the short-range interactions are suppressed by the GIM mechanism. The inclusive branching fraction~(BF) arising solely from the short-range penguin diagrams is only at the order of magnitude of $10^{-8}$~\cite{Fajfer:1998dv}, which is much smaller than the contributions from long-range effects. 
The long-range non-perturbative processes can enhance the BF up to $10^{-4}$~\cite{Altmannshofer:2022hfs}. 
The decay diagram for $D^+_s\to\gamma K^*(892)^{+}$ via a long-range interaction is shown in Figure~\ref{fig:feynman1}.
The decays $D \to VV^{'}$, where $V$ and $V^{'}$ represent vector mesons, have large BFs due to long-range dynamics such as final-state interactions (FSIs) and $V^{(')}$ helicity contributions.
Via vector meson dominance~(VMD), the vector meson $V'$ can convert to a real photon, yielding a non-negligible long-range contribution to the radiative decay $D \to \gamma V$. This VMD connection means $D \to \gamma V$ may receive enhanced contributions from the same long-range dynamics boosting $D \to VV'$ BFs~\cite{Cao:2023gfv,Cao:2023csx}.


Various models have predicted the BF of the Cabibbo-suppressed process $D_s^+ \to \gamma K^*(892)^+$, with values ranging from ${\cal O}(10^{-5})$ to ${\cal O}(10^{-4})$~\cite{deBoer:2017que,Lyon:2012fk,Fajfer:1998dv,Fajfer:1997bh,Burdman:1995te}. These predictions are based on the Standard Model and employ different theoretical frameworks, including hard spectator interaction~(HSI) and weak annihilation~(WA)~\cite{deBoer:2017que}, light-cone sum rules~(LCSR)~\cite{Lyon:2012fk}, hybrid framework~\cite{Fajfer:1998dv,Fajfer:1997bh}, 
VMD~\cite{Burdman:1995te}. 
Experimentally, many radiative decays of charmed mesons have been studied. 
The BESIII Collaboration reported the upper limits (ULs) on the BF of $D^+_s \to \gamma\rho(770)^+$~\cite{BESIII:2024thr} and $D^+ \to \gamma\rho(770)^+$ and $D^+ \to \gamma K^*(892)^+$~\cite{BESIII:2024zfv}. 
The BaBar Collaboration measured the BFs of $D^0 \to \gamma\bar{K}^*(892)^0$ and $D^0 \to \gamma\phi(1020)$~\cite{BaBar:2008kjd}. 
The CLEO Collaboration obtained the ULs of $D^0 \to \gamma\bar{K}^*(892)^0$, $D^0 \to \gamma\rho(770)^0$, $D^0 \to \gamma\omega(782)$, and $D^0 \to \gamma\phi(1020)$~\cite{CLEO:1998mtp}.
The Belle Collaboration provided the BFs of $D^0 \to \gamma\bar{K}^*(892)^0$ and $D^0 \to \gamma\rho(770)^0$~\cite{Belle:2016mtj} and $D^0 \to \gamma\phi(1020)$~\cite{Belle:2016mtj,Belle:2003vsx}.
Additionally, a sensitivity study of $D_s^+ \to \gamma K^*(892)^+$ and $D_s^+ \to \gamma \rho(770)^+$ has been performed by the Belle Collaboration~\cite{Ipsita:2024wpo}.
 However, the BF of the radiative decay $D^+_s\to\gamma K^*(892)^{+}$ has not yet been measured experimentally. Overall, the experimentally measured BFs for radiative charm decays range from $10^{-5}$ to $10^{-4}$, consistent with theoretical predictions.

In this paper, we report the first search for the radiative decay $D_s^+\to\gamma K^*(892)^+$ using $e^+e^-$ collision data corresponding to an integrated luminosity of $7.33\,\mathrm{fb}^{-1}$ collected with the BESIII detector at center-of-mass energies ($E_{\mathrm{cm}}$) between $4.128$ and $4.226\,\mathrm{GeV}$. 
The \( K^*(892)^+ \) is reconstructed via its decays to \( K^+ \pi^0 \) or \( K_S^0 \pi^+ \). Charge-conjugated modes are implicitly considered throughout this paper.

\section{Detector and data sets}
\label{sec:detector_dataset}
The BESIII detector~\cite{BESIII:2009fln} records symmetric $e^+e^-$ collisions provided by the BEPCII storage ring~\cite{Yu:2016cof} in the center-of-mass energy range from 1.84 to 4.95 GeV, with a peak luminosity of $1.1\times 10^{33}$ cm$^{-2}$s$^{-1}$ achieved at $\sqrt s$ = 3.773 GeV.
BESIII has collected large data samples in this energy region~\cite{BESIII:2020nme,Zhang:2022bdc,Lu:2020imt}.
The cylindrical core of the BESIII detector covers 93\% of the full solid angle and consists of a helium-based multilayer drift chamber~(MDC), 
a plastic scintillator time-of-flight system~(TOF), and a CsI(Tl) electromagnetic calorimeter~(EMC), 
which are all enclosed in a superconducting solenoidal magnet providing a 1.0~T magnetic field. 
The solenoid is supported by an octagonal flux-return yoke with resistive plate counter muon identification modules interleaved with steel.
The charged-particle momentum resolution at 1~GeV/$c$ is $0.5\%$, and the ${\rm d}E/{\rm d}{x}$ resolution is $6\%$ for electrons from Bhabha scattering. 
The EMC measures photon energies with a resolution of $2.5\%$ ($5\%$) at 1~GeV in the barrel (end-cap) region.
The time resolution in the TOF barrel region is 68~ps, while that in the end-cap region is 110~ps. 
The end-cap TOF system was upgraded in 2015 using multi-gap resistive plate chamber technology, providing a time resolution of 60~ps~\cite{Li:2017jpg, Guo:2017sjt, Cao:2020ibk}. About 83$\%$ of the data in this analysis benefits from the upgrade.

To reduce potential systematic uncertainties associated with different detector conditions across years,
the data are grouped into four samples according to running period and conditions: $E_{\rm cm} =$ 4.128 and 4.157 GeV~(I), 4.178 GeV~(II),  four energies from 4.189 to 4.219 GeV~(III), and 4.226 GeV~(IV).
Integrated luminosities~\cite{BESIII:2022dxl} at each energy are given in Table~\ref{energe}.
\begin{table}[htpb]
\centering
\begin{tabular}{c c c c}
\hline
$E_{\rm cm}$ (GeV)~\cite{BESIII:2020eyu, BESIII:2015zbz} &$\mathcal{L}_{\rm int}$ (pb$^{-1}$) & $M_{\rm rec}$ (GeV/$c^2$) &Group\\
\hline 
4.128 &401.5   &[2.060, 2.150] & \multirow{2}{*}{\centering I}\\
	 4.157 &408.7   &[2.054, 2.170] &\\
  \hline
	 4.178 &$3189.0\pm0.2\pm31.9$  &[2.050, 2.180] &II\\
  \hline
	 4.189 &$570.0\pm0.1\pm2.2$ &[2.048, 2.190] & \multirow{4}{*}{\centering III}\\
	 4.199 &$526.0\pm0.1\pm2.1$ &[2.046, 2.200] &\\
	 4.209 &$572.1\pm0.1\pm1.8$ &[2.044, 2.210] &\\
	 4.219 &$569.2\pm0.1\pm1.8$ &[2.042, 2.220] &\\
  \hline
	 4.226 &$1100.9\pm0.1\pm7.0$ &[2.040, 2.220] &IV\\
\hline
\end{tabular}
\caption{\label{energe} The integrated luminosities ($\mathcal{L}_{\rm int}$) and the requirements on the $D^-_s$ recoil mass ($M_{\rm rec}$) for various center-of-mass energies.
	 The definition of $M_{\rm rec}$ is given in eq.~(\ref{eq:mrec}).
   The first and second uncertainties for $\mathcal{L}_{\rm int}$ are statistical and systematic, respectively. The integrated luminosities for the two data samples at $E_{\rm cm} = 4.128$ GeV and $E_{\rm cm} = 4.157$ GeV are estimated by using online monitoring information.}
\end{table}

Monte-Carlo (MC) samples simulated with a {\sc geant4}-based~\cite{GEANT4:2002zbu} package, including the full BESIII geometry and response, are used to determine efficiencies and estimate backgrounds.  The simulation models the beam energy spread and initial state radiation (ISR) in the $e^+e^-$ annihilations with the generator {\sc kkmc}~\cite{Jadach:2000ir, Jadach:1999vf}. 
Inclusive MC samples of 40 times the size of the data samples are used to simulate the background contributions. The inclusive MC samples include the production of open charm processes, the ISR production of vector charmonium(-like) states, and the continuum processes incorporated in {\sc kkmc}. All particle decays are modeled with {\sc evtgen}~\cite{Lange:2001uf, Ping:2008zz} using BFs either reported by the PDG~\cite{PDG}, 
when available, or otherwise estimated with {\sc lundcharm}~\cite{Chen:2000tv, Yang:2014vra}. 
Final state radiation from charged particles is incorporated using {\sc photos}~\cite{Richter-Was:1992hxq}. 

Signal MC samples of 1.5 million events each are generated for the two $K^*(892)^+$ decay chains: $D_s^+\to\gamma K^*(892)^+$, $K^*(892)^+\to K^+\pi^0$, $\pi^0\to\gamma\gamma$ and $D_s^+\to\gamma K^*(892)^+$, $K^*(892)^+\to K_S^0\pi^+$, $K_S^0\to\pi^+\pi^-$.
The signal decay $D_s^+ \to \gamma K^*(892)^+$ is modelled with helicity amplitudes~\cite{Lange:2001uf, Ping:2008zz} giving in an angular distribution of $1 - \cos^2\theta_{\mathrm{H}}$~\cite{Belle:2016mtj, BaBar:2008kjd}.
Here, $\theta_{\mathrm{H}}$ is the helicity angle between the $K^+$ ($\pi^+$) momentum in the $K^*(892)^+$ rest frame and the $K^*(892)^+$ flight direction in the $D_s^+$ rest frame.  $K^*(892)^+$ decays are simulated with the VSS model~\cite{Lange:2001uf, Ping:2008zz}, which describes the decay of a vector meson into a pair of scalar mesons.

\section{Analysis strategy}
In $e^+e^-$ collisions at $E_{\mathrm{cm}}$ between $4.128$ and $4.226$ GeV, $D_s^{\pm}$ mesons are produced mainly via the process $e^+e^- \to D_s^{* \pm}D_s^{\mp}$, with the  $D_s^{* \pm}$ subsequently decaying into 
$\gamma(\pi^0)D_s^+D_s^-$. We perform this analysis by using the double-tag (DT) method pioneered by  MARK-III~\cite{MARK-III:1985hbd, Ke:2023qzc}. 
A $D_s^-$ meson fully reconstructed via one of its hadronic decay modes is referred to as a single-tag (ST) $D_s^-$ meson.
In the system recoiling against the ST $D_s^-$ meson, we search for the radiative decay $D_s^+\to\gamma K^*(892)^+$ through the two channels, $K^*(892)^+\to K^+\pi^0$ and $K^*(892)^+\to K_S^0\pi^+$. The combination of a ST $D_s^-$ candidate and a signal $D_s^+\to\gamma K^*(892)^+$ candidate is defined as a DT candidate.

The BF of the signal decay is determined by
\begin{eqnarray}\begin{aligned}
\mathcal{B}(D_{s}^{+} \to \gamma K^*(892)^+)
	 =
	\frac{N^{\rm DT}_{\rm total}}{\mathcal{B}_{\rm sub}\sum_{\alpha,i}{N^{\rm ST}_{\alpha,i}}\epsilon^{\rm DT}_{\alpha,i}/\epsilon^{\rm ST}_{\alpha,i}},
	\label{abs:bf}
\end{aligned}\end{eqnarray}
where $N_{\rm total}^{\rm DT}$ is the sum of the DT yields for each of the $K^*(892)^+\to K^+\pi^0$ and $K^*(892)^+\to K_S^0\pi^+$ channels, with each channel summed over all four sample groups.
$\epsilon^{\rm ST}_{\alpha,i}$ is the ST efficiency to reconstruct the ST mode $i$ in the sample group $\alpha$, $\epsilon^{\rm DT}_{\alpha,i}$ is the DT efficiency for reconstructing the ST mode $i$ and the signal decay mode in the sample group $\alpha$.
$\mathcal{B}_{\mathrm{sub}}$ is the product of BFs for $K^*(892)^+\to K^+\pi^0$, $\pi^0\to\gamma\gamma$ or $K^*(892)^+\to K_S^0\pi^+$, $K_S^0\to\pi^+\pi^-$.

\section{Single tag selection}
\label{ST-selection}
Charged tracks detected in the MDC are required to be within a polar angle ($\theta$) range of $|\rm{cos\theta}|<0.93$, where $\theta$ is defined with respect to the $z$-axis, which is the symmetry axis of the MDC. 
For charged tracks not originating from $K_S^0$ decays, the distance of closest approach to the interaction point~(IP) is required to be less than 10~cm along the $z$-axis, $|V_{z}|$, and less than 1~cm in the transverse plane, $|V_{xy}|$.
Particle identification~(PID) for charged tracks combines the measurements of d$E$/d$x$ in the MDC and the time of flight in the TOF to form likelihoods $\mathcal{L}(h)~(h=K,\pi)$ for each hadron hypothesis.
Charged kaons and pions are identified by comparing the likelihoods for the two hypotheses, $\mathcal{L}(K)>\mathcal{L}(\pi)$ and $\mathcal{L}(\pi)>\mathcal{L}(K)$.

$K^0_S$ candidates are reconstructed with two oppositely charged tracks that satisfy $|V_{z}|<$ 20~cm, assigned as $\pi^+\pi^-$ and without PID imposed.
In order to get a better mass resolution of $K_S^0$, the $\pi^+\pi^-$ pairs are constrained to have a common vertex and the $\chi^2$ is required less than 100 for the vertex fit.
The $K^0_S$ candidates are required to have the invariant mass~($M_{\pi^{+}\pi^{-}}$) within [0.487,0.511] GeV/$c^2$.
The decay length of the $K_S^0$ candidate is required to be greater than twice the vertex resolution away from the IP. 

Photon candidates are identified using showers in the EMC. 
To suppress backgrounds from electronic noise or bremsstrahlung, any candidate shower is required to start within $[0,700]\ \mathrm{ns}$ from the event start time.
The deposited energy of each shower must be more than 25~MeV in the barrel region~($|\!\cos \theta|< 0.80$) and more than 50~MeV in the end-cap region~($0.86 <|\!\cos \theta|< 0.92$). 
To exclude showers that originate from charged tracks,
the opening angle subtended by the EMC shower and the position of the closest charged track at the EMC
must be greater than $10^\circ$ as measured from the IP. 

$\pi^0$ candidates are reconstructed through the $\pi^0\to \gamma\gamma$ decay, with at least one photon being detected in the barrel region. 
The invariant mass of the photon pair must be in the range of [$0.115, 0.150$]~GeV/$c^{2}$, approximately three times the mass resolution around the known $\pi^0$ mass~\cite{PDG}. 
To improve the mass resolution, a kinematic fit is performed on the selected photon pair, constraining the $\gamma\gamma$ invariant mass to the known $\pi^{0}$ mass. The $\chi^2$ is required to be less than 30. 

To veto soft pions from $D^{*+}$ decays, the minimum momentum of any pion which is not from $K_S^0$ is required to be greater than $0.1\ \mathrm{GeV}/c$.
To avoid double-counting an event in both the $D_s^-\to K_S^0K^-$ and $D_s^-\to K^-\pi^{-}\pi^{+}$ ST modes, $M_{\pi^{+}\pi^{-}}$ is required to be outside the mass range $[0.487, 0.511]$ GeV$/c^{2}$ for the $D_s^-\to K^-\pi^{-}\pi^{+}$ mode. 

\begin{table}[tp]
\centering
\begin{tabular}{lc}
\hline
Tag mode                                     & Mass window (GeV/$c^{2}$) \\
\hline 
$D_{s}^{-} \to K_{S}^{0}K^{-}$               & [1.948, 1.991]            \\
        $D_{s}^{-} \to K^{+}K^{-}\pi^{-}$            & [1.950, 1.986]            \\
        $D_{s}^{-} \to K_{S}^{0}K^{+}\pi^{-}\pi^{-}$ & [1.953, 1.983]            \\
\hline
\end{tabular}
\caption{\label{tab:tag-cut} Requirements on $M_{\rm tag}$ for different tag modes.}
\end{table}

\begin{figure}[htbp]
    \centering   \includegraphics[width=0.55\linewidth]{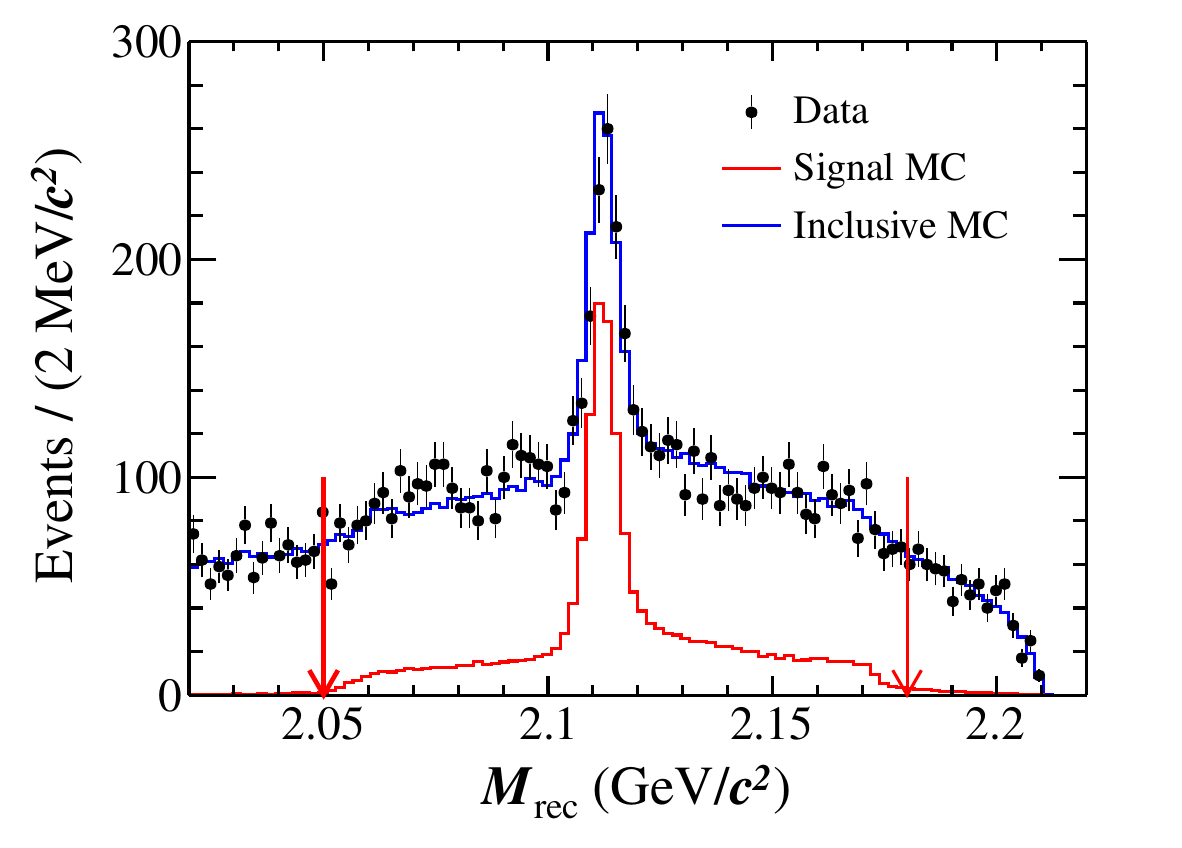}
    \caption{$M_{\rm rec}$ distribution for $D_s^+ \to K^+K^-\pi^+$ at $E_{\text{cm}} = 4.178$ GeV. The dots with error bars represent the data points. 
    The red solid line represents the signal shape scaled to the data.
     The blue histogram represents inclusive MC samples normalized to the data.
     The red arrows indicate the $M_{\rm rec}$ cut region.}
    \label{mrec_tag}
\end{figure}

Candidates for the $D_s^-$ meson are reconstructed through the three ST modes, chosen following the optimisation discussed in Section 5. For each tag mode, the corresponding mass windows on the invariant mass of the reconstructed ST $D_s^-$ ($M_{\mathrm{tag}}$) are listed in Table~\ref{tab:tag-cut}.
The recoiling mass ($M_{\rm rec}$) against the ST $D_s^-$ is evaluated, and events with $M_{\rm rec}$ within the mass windows  specified in Table~\ref{energe} are retained for further analysis.
These $M_{\rm rec}$ ranges are determined by ensuring the tag efficiencies remain roughly constant across the different $E_{\rm cm}$ of the data samples. 
For directly produced $D_s$ from $e^+e^- \to D_s^{*+}D_s^-$ or $D_s^{*-}D_s^+$, $M_{\rm rec}$ peaks at the $D_s^*$ mass of $2.112\ \mathrm{GeV}/c^2$;
for $D_s$ candidates originating from $D_s^*$ decays (indirect $D_s$), $M_{\rm rec}$ is distributed around this peak. 
As an example, Figure~\ref{mrec_tag} shows the $M_{\rm rec}$ distribution for the tag mode $D_s^- \to K^+ K^- \pi^-$ at $E_{\rm cm} = 4.178$~GeV.
The recoiling mass $M_{\rm rec}$ is defined as
\begin{eqnarray}
\begin{aligned}
	\begin{array}{lr}
	M_{\rm rec} = \sqrt{\left(E_{\rm cm}/c^2 - \sqrt{|\vec{p}_{D_{s}^-}|^{2}/c^{2}+m_{D_{s}^-}^{2}}\right)^{2} - |\vec{p}_{D_{s}^-}|^{2}/c^{2}} \; , \label{eq:mrec}
		\end{array}
\end{aligned}\end{eqnarray}
where $E_{\rm cm}$ is the center-of-mass energy of the data sample, $\vec{p}_{D_{s}^-}$ is the three-momentum of the ST $D_{s}^{-}$ candidate in the $e^+e^-$ center-of-mass frame, and $m_{D_{s}^-}$ is the known  $D_{s}^{-}$ mass~\cite{PDG}.

\begin{figure*}[htp]
\begin{center}
    \includegraphics[width=13.cm]{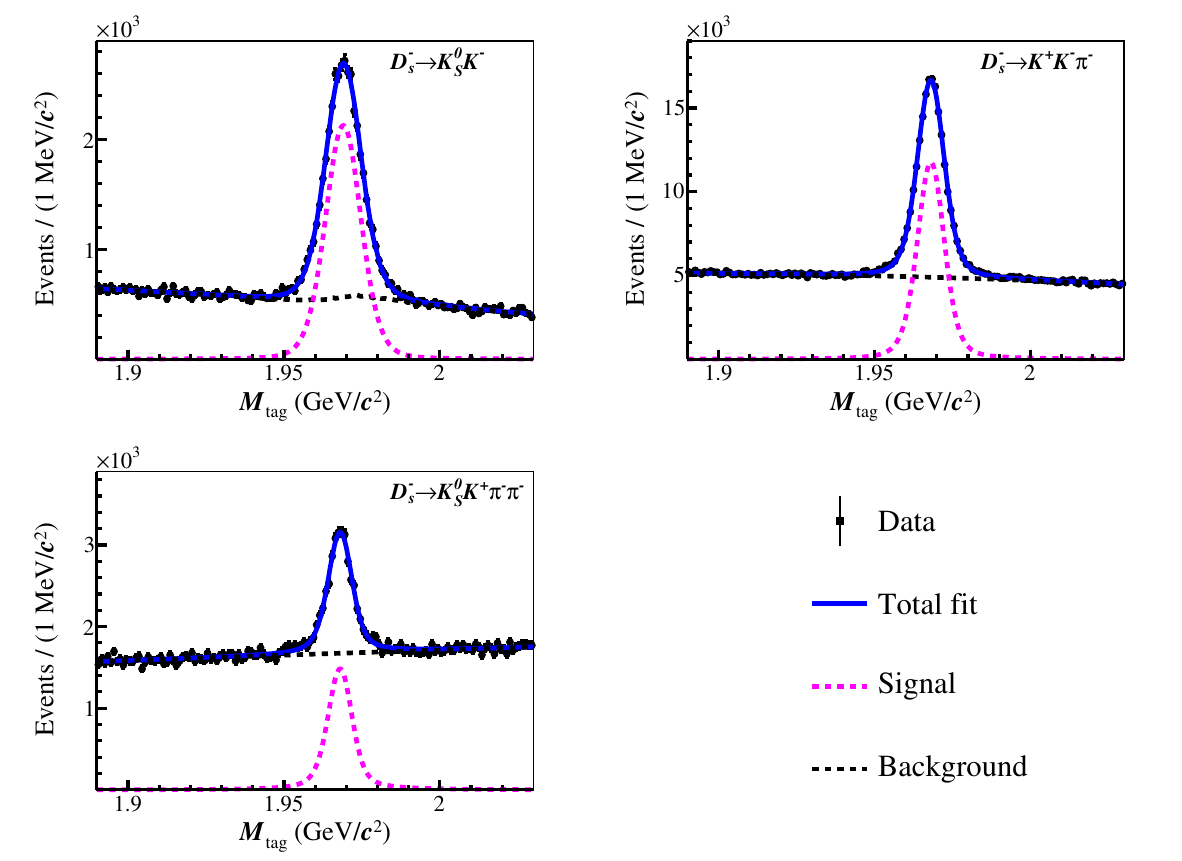}
\caption{Fits to the $M_{\rm tag}$ distributions of the ST candidates from the data sample at $E_{\rm cm} = 4.178$~GeV. 
	The black points with error bars are data, the blue solid lines are the total fit results, the magenta dashed lines show the signal component and the black dashed lines show the background component. }
\label{fit:Mass-data-Ds_4180}
\end{center}
\end{figure*}

For an event with multiple candidates for a specific tag mode per charge, 
only the one with $M_{\rm rec}$ closest to the known $D_s^{*+}$ mass~\cite{PDG} is chosen. 
The yields of the ST modes, derived from binned maximum-likelihood 
fits to $M_{\mathrm{tag}}$
for each individual ST decay mode, are listed in Table~\ref{ST-yield}.
As an illustrative example, the fits to the accepted ST candidates from the data sample at $E_{\rm cm} = 4.178$~GeV are shown in 
Figure~\ref{fit:Mass-data-Ds_4180}.
In the fits, the signal is modeled by an MC-simulated shape convolved with a Gaussian function that accounts for the data-MC difference in detector resolution.
The background is described by a second-order Chebyshev polynomial. 
For the tag mode $D_{s}^{-} \to K_{S}^{0} K^-$, the peaking background originating from $D^{-} \to K_{S}^{0} \pi^-$ is considered.
	The shape of this background is taken from the inclusive MC samples and included in the fit with a free yield parameter. 
	
The ST efficiencies in Eq.~(\ref{abs:bf}) are estimated using the inclusive MC samples.
The same selection criteria and fitting approach as applied to the ST data are used to analyse the inclusive MC samples. 
The number of observed ST events is extracted from fits to the $M_{\rm tag}$ distributions. 
The ST efficiency is calculated as the ratio of the observed ST events and the generated ST events in the inclusive MC samples, as shown in  Table~\ref{ST-eff}. 
To determine the ST efficiency for a specific grouped dataset~(I and III), the yields at different energy points are averaged based on respective integrated luminosities and cross sections.

\begin{table*}[tp]
	\centering
  \begin{tabular}{l r@{$\pm$}l r@{$\pm$}l r@{$\pm$}l r@{$\pm$}l }
  \hline
    Tag mode   &  \multicolumn{2}{c}{$N^{\text{ST}}_{\rm I}$}  
    &\multicolumn{2}{c}{$N^{\text{ST}}_{\rm II}$}   
		&\multicolumn{2}{c}{$N^{\text{ST}}_{\rm III}$}     
		&\multicolumn{2}{c}{$N^{\text{ST}}_{\rm IV}$}   \\
  \hline
    $D^-_s\to K^0_SK^-$          &6728  &144  &31949  &314  &19960 &270  &6837  &163\\
    $D^-_s\to K^+K^-\pi^-$       &27670 &280  &137138 &614  &86918 &525  &29544 &335  \\
    $D^-_s\to K_S^0K^+\pi^-\pi^-$&2983  &129  &15705  &288  &9783  &247  &3380  &174\\
    \hline
  \end{tabular}\caption{The ST yields~($N^{\text{ST}}_{\alpha, i}$) for the data samples collected at $E_{\rm cm} =$ 4.128 and 4.157~GeV (I), 4.178~GeV (II), from 4.189 to 4.219~GeV (III), 4.226~GeV (IV). The uncertainties are statistical.}
  \label{ST-yield}
\end{table*}

\begin{table*}[tp]
	\centering
  \begin{tabular}{l r@{$\pm$}l r@{$\pm$}l r@{$\pm$}l r@{$\pm$}l}
  \hline
    Tag mode   &\multicolumn{2}{c}{$\epsilon^{\rm ST}_{\rm I} (\%)$} 
    &\multicolumn{2}{c}{$\epsilon^{\rm ST}_{\rm II} (\%)$}
		&\multicolumn{2}{c}{$\epsilon^{\rm ST}_{\rm III} (\%)$}  
		&\multicolumn{2}{c}{$\epsilon^{\rm ST}_{\rm IV} (\%)$}\\
  \hline
    $D^-_s\to K^0_SK^-$          &47.64 &0.16 &47.39 &0.07 &47.23 &0.09  &47.95 &0.16\\
    $D^-_s\to K^+K^-\pi^-$       &40.37 &0.07 &39.47 &0.03 &39.33 &0.04  &39.78 &0.07\\
    $D^-_s\to K_S^0K^+\pi^-\pi^-$&21.30 &0.14 &21.85 &0.06 &21.66 &0.08  &22.27 &0.16\\
    \hline
  \end{tabular}
  \caption{The ST efficiencies~($\epsilon^{\rm ST}_{\rm {\alpha, i}}$) for the data samples collected at $E_{\rm cm} =$ 4.128 and 4.157~GeV (I), 4.178~GeV (II), from 4.189 to 4.219~GeV (III), 4.226~GeV (IV). The uncertainties are statistical.}
  \label{ST-eff}
\end{table*}

\section{Double tag selection}
In the system recoiling against tagged $D_s^-$ candidates, we search for the signal process $D^+_s\to \gamma K^*(892)^+$, with $K^*(892)^+\to K^+\pi^0$ or $K_S^0\pi^+$. 
Selection criteria for $\gamma$, $K^{\pm}$, $\pi^{\pm}$, $K_S^0$ and $\pi^0$ are the same as those for ST.
The radiative photon is taken to be the most energetic photon in the event.   
If there are multiple DT candidates in an event, the candidate with the average mass $(M_{\rm sig}+M_{\rm tag})/2$ closest to the known $D_{s}^{\pm}$ mass is retained, where $M_{\rm sig}$ represents the invariant mass of the accepted signal $D^+_s$. 

To select the $K^*(892)^+$ and suppress combinatorial background, we require the signal candidates to satisfy $M_{K^+\pi^0} \in$ (0.83, 0.94) GeV/$c^2$ for $K^*(892)^+\to K^+\pi^0$ channel and $M_{K_S^0\pi^+} \in$ (0.83, 0.94) GeV/$c^2$ for $K^*(892)^+\to K_S^0\pi^+$ channel. 
To further effectively suppress the background involving $\pi^0$ or $\eta$ mesons, we require the energy of radiative photon $E_{\gamma}>0.55$ GeV/$c^2$, resulting in a negligible signal efficiency loss of $0.22\%$ for $K^*(892)^+\to K^+\pi^0$ channel and $0.24\%$ for $K^*(892)^+\to K_S^0\pi^+$ channel.

For both $K^*(892)^+$ decay channels, 
to suppress backgrounds involving $\pi^0$ and $\eta$ mesons,
we veto candidates where the invariant mass of $M_{\gamma\gamma_{\rm extra}}$ falls into the $\pi^0$ mass window [0.115, 0.150] GeV/$c^2$ or the $\eta$ mass window [0.50, 0.57] GeV/$c^2$, in which $\gamma_{\rm extra}$ is an extra photon not used in either the ST or signal sides. 
Additionally, since photons from the process $D_s^{*\pm} \to \gamma D_s^{\pm}$ typically carry an energy of around 0.15 GeV, we also apply an energy requirement $E_{\gamma_{\rm extra}} > 0.2$ GeV.
This $M_{\gamma\gamma_{\rm extra}}$ veto has a rejection efficiency of $73\%$ ($77\%$) with a signal loss of $32\%$ ($28\%$) for the $K^+\pi^0$ ($K_S^0\pi^+$) channel. 

For the $K^+\pi^0$ channel specifically, to further suppress background events from $D^+_s \to K^+ \eta$, $\eta \to \gamma\gamma$, we impose an additional requirement $M_{\gamma\gamma_{\mathrm{h}}} > 0.62$ GeV/$c^2$, ensuring the invariant mass is above the nominal $\eta$ mass. Here, $\gamma_{\mathrm{h}}$ is the higher-energy photon from the $\pi^0$ decay on the signal side.



The selection windows for $M_{K^+\pi^0}$, $M_{K_S^0\pi^+}$, and $M_{\gamma\gamma_{\text{h}}}$ are optimised by maximising the Punzi figure-of-merit $\frac{\epsilon}{1.5+\sqrt{B}}$~\cite{Punzi:2020fsv}, where $\epsilon$ is the signal efficiency based on the signal MC sample, 
and $B$ is the background yield obtained from the inclusive MC samples and normalized to the data.

Based on these optimized selection criteria, we further determine the combination of tag modes using MC simulation.
We perform the optimization
by maximizing the figure-of-merit $S/\sqrt{S + B}$, where $S$ is the expected signal yield based on the predicted BF of $D_{s}^{+} \to \gamma K^*(892)^+$~\cite{deBoer:2017que} and $B$ is the background yield obtained from the inclusive MC samples and normalized to the data. 
Finally, we select ST $D_s^-$ candidates by using three hadronic decay modes: $D_{s}^{-}\to K_{S}^{0}K^{-}$, $D_{s}^{-}\to K^{+}K^{-}\pi^{-}$, and $D_{s}^{-}\to K_S^0K^{+}\pi^-\pi^-$.


After applying the above selection criteria, 
the background is divided into two categories:
$D^{*\pm}_{s}D_s^{\mp}$ processes~(denoted as $B_{\rm D_s^{+}}$) and non-$D^{*\pm}_{s}D_s^{\mp}$ backgrounds~(denoted as ${B}_{\rm{other}}$). The dominant $D^{*\pm}_{s}D_s^{\mp}$ background is $D_s^+ \to \pi^+ \pi^0 \eta$ decay, while the non-$D^{*\pm}_{s}D_s^{\mp}$ background is dominated by $q\bar{q}$ continuum events.
The DT efficiencies are calculated as the ratio of the number of observed DT events to the number of generated DT events in the signal MC samples. The resulting efficiencies for each tag mode and sample group are listed in Table~\ref{abs:dteff}.

\section{Result}
\label{BFSelection} 
After the above requirements, 
the two-dimensional~(2D) distributions of $M_{\mathrm{sig}}$ versus $\cos\theta_{\mathrm{H}}$ of $K^+ \pi^0 (K_S^0\pi^+)$ for the data are shown in Figure~\ref{2d-dali2}. 
Based on the analysis of inclusive MC samples, the higher background in the $K_S^0\pi^+$ channel is due to larger combinatorial backgrounds from $q\bar{q}$ events and $D_s^+ \to \eta\pi^+\pi^0$ decays with $\eta\to\pi^+\pi^-\pi^0$ and $\eta\to\gamma\gamma$, while the $K^+\pi^0$ channel benefits from an additional $M_{\gamma\gamma_{\mathrm{h}}}$ cut that suppresses backgrounds more effectively.

We perform a simultaneous fit to search for the radiative decay $D_s^+\to\gamma K^*(892)^+$ via $K^*(892)^+\to K^+\pi^0$ and $K^*(892)^+\to K_S^0\pi^+$. In the fit, the BF is a common parameter shared by both channels.
The same set of tag modes is used for both $K^*(892)^+$ decay channels, and the data samples from all tag modes and all $E_{\rm cm}$ sample groups are combined according to their efficiencies into a total sample for the fit.
The {$N_{\rm Total}^{\text{DT}}$ for the two decay channels can be derived from the eq.~(\ref{abs:bf}), respectively.
The isospin relation ${\cal B}(K^*(892)^+\to K_S^0\pi^+)/{\cal B}(K^*(892)^+\to K^+\pi^0)=2$ is imposed.

Projections of the fit results are shown in Figure~\ref{2dfit3}.
The BF is obtained from a 2D unbinned maximum likelihood fit on the $M_{\rm sig}$ versus cos$\theta_{\rm H}$ distribution for $D_s^+\to\gamma K^*(892)^+$. The signal cos$\theta_{\rm H}$ distribution follows $1 - \cos^2\theta_{\rm H}$ due to angular momentum conservation, which distinguishes it from background.
The signal shape is described by a MC-simulated 2D probability density function (PDF) convolved with a 2D Gaussian function to account for the resolution difference between the data
and the signal MC.
The shapes of both $B_{D_s^+}$ and $B_{\rm other}$ are derived from inclusive MC samples, and their yields are allowed to float in the fit.
The fitted signal yields are 
$0.0\pm1.8$ for the $K^*(892)^+\to K^+\pi^0$ channel and 
$0.1\pm1.1$ for the $K^*(892)^+\to K_S^0\pi^+$ channel.

\begin{figure}[tp!]
\centering	
\includegraphics[width=7.2cm]
{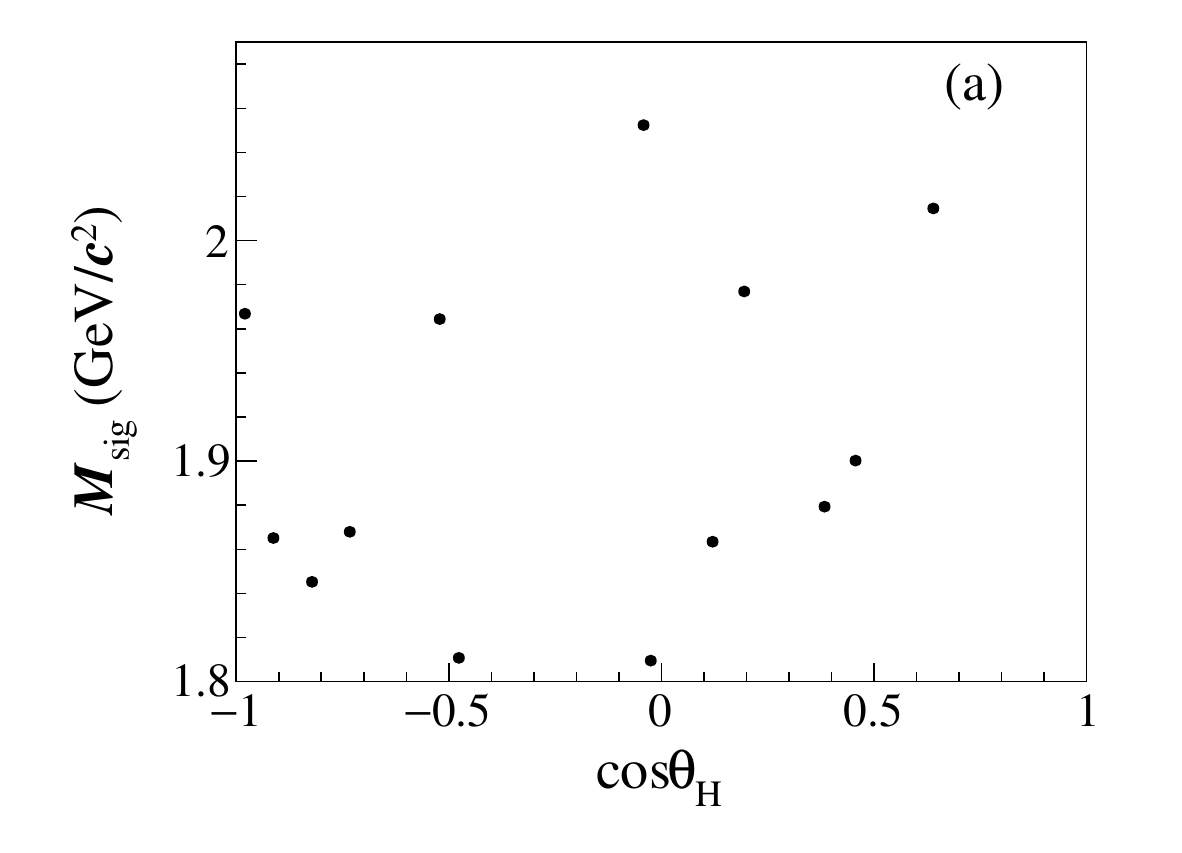}
\includegraphics[width=7.2cm]
{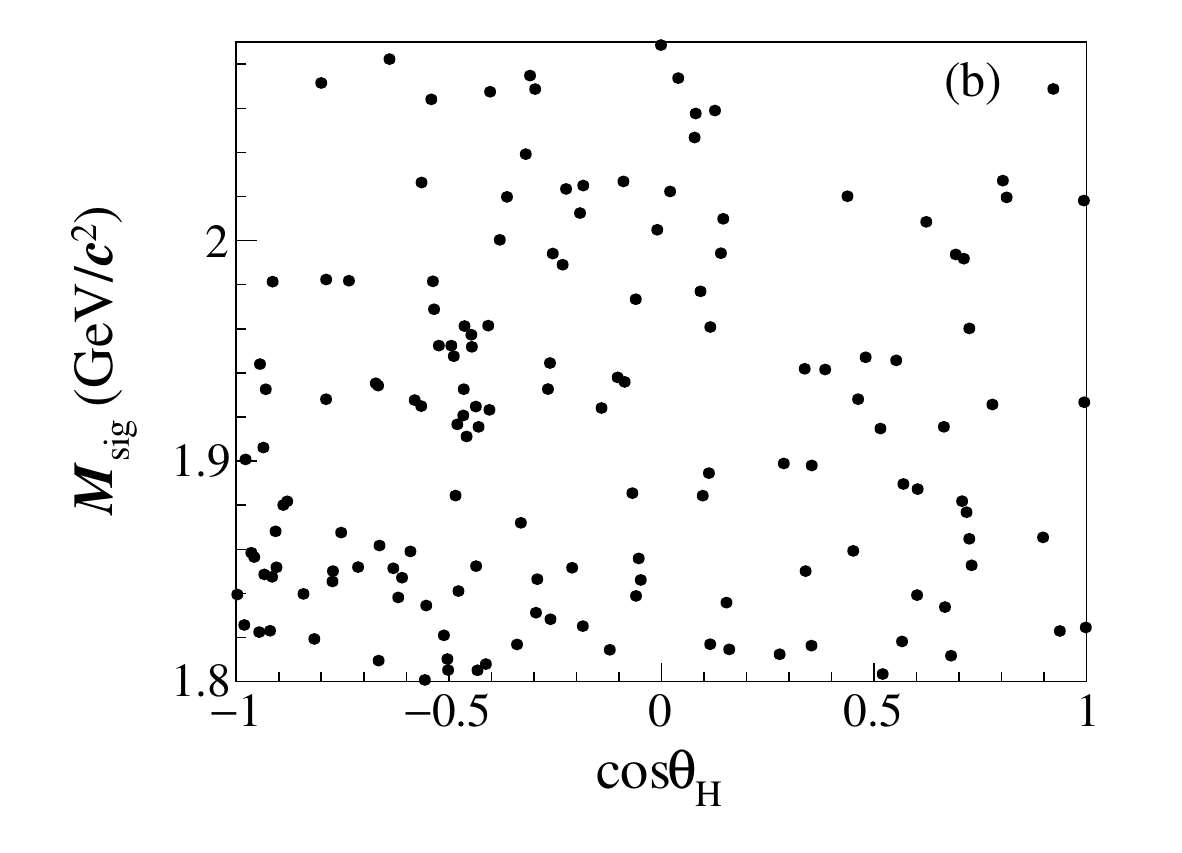}\\
\includegraphics[width=7.2cm]
{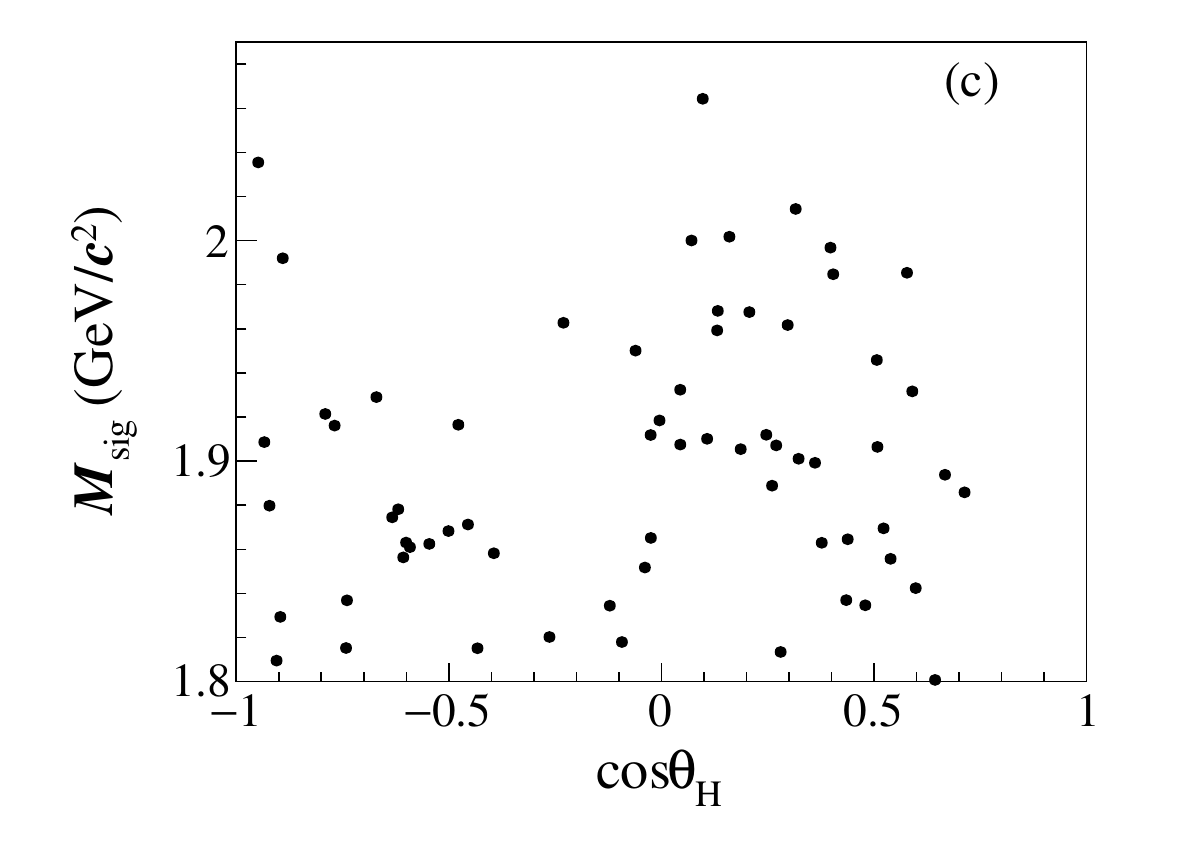}
\includegraphics[width=7.2cm]
{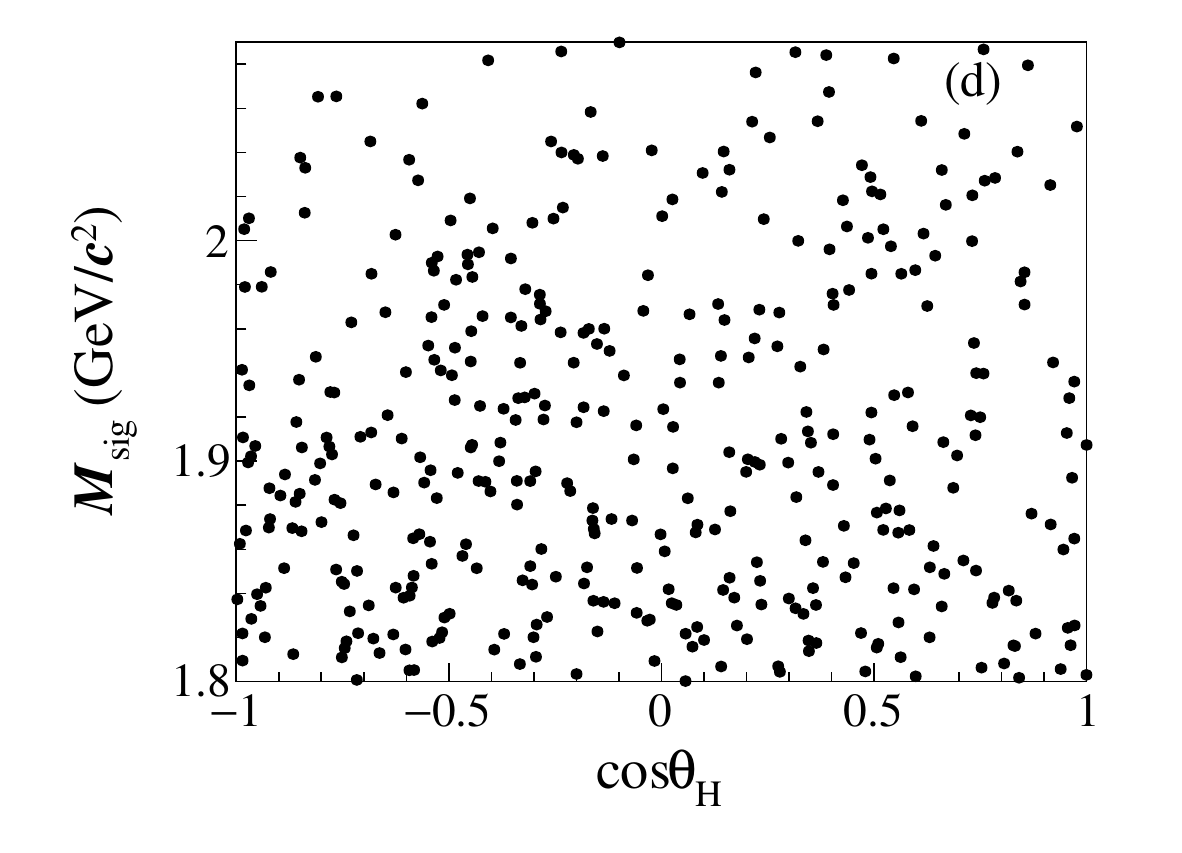}
\caption{Distributions of $M_{D_s^+}$ versus cos$\theta_{\rm H}$ of the DT candidate events for $D_s^+\to\gamma K^*(892)^+$ via $K^*(892)^+\to K^+\pi^0$~((a) and (c)) and $K^*(892)^+\to K_S^0\pi^+$~((b) and (d)) in data~((a) and (b)) and inclusive MC~((c) and (d)), where inclusive MC samples are five times the size of the data sample.}
\label{2d-dali2}
\end{figure}

\begin{figure}[htbp]
          \centering
          \begin{minipage}{7.3cm}
    \includegraphics[width=7.3cm]{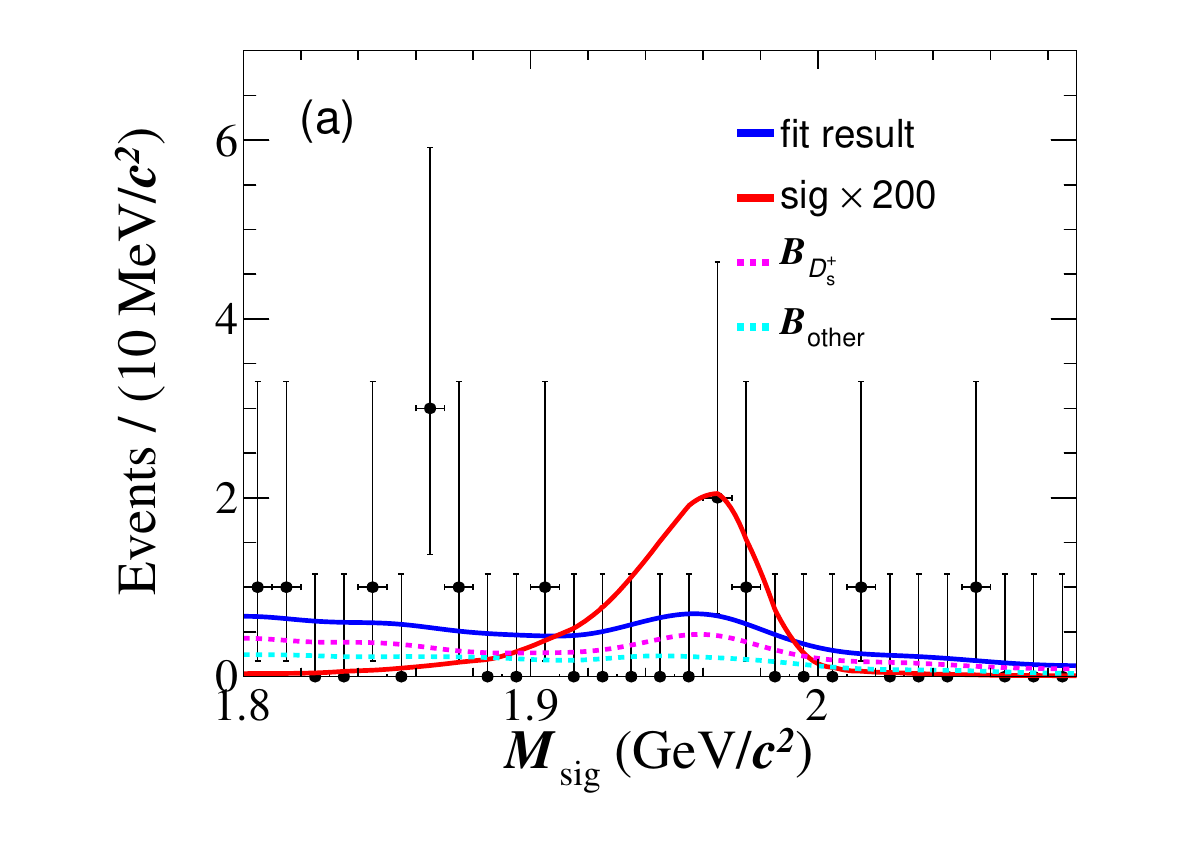}
    \end{minipage}   
\begin{minipage}{7.3cm}
    \includegraphics[width=7.3cm]{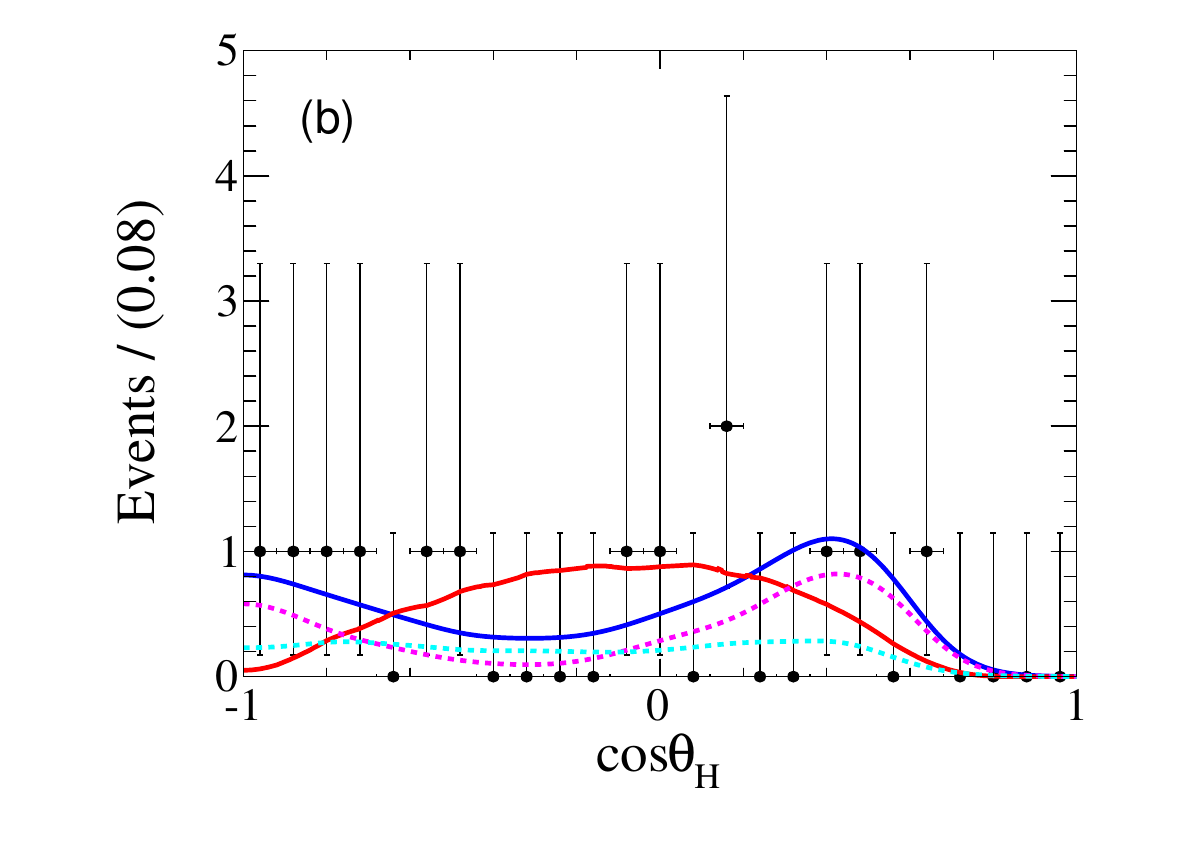}
  \end{minipage}
  \hspace*{0.0001cm}
  \begin{minipage}{7.3cm}
 \includegraphics[width=7.3cm]{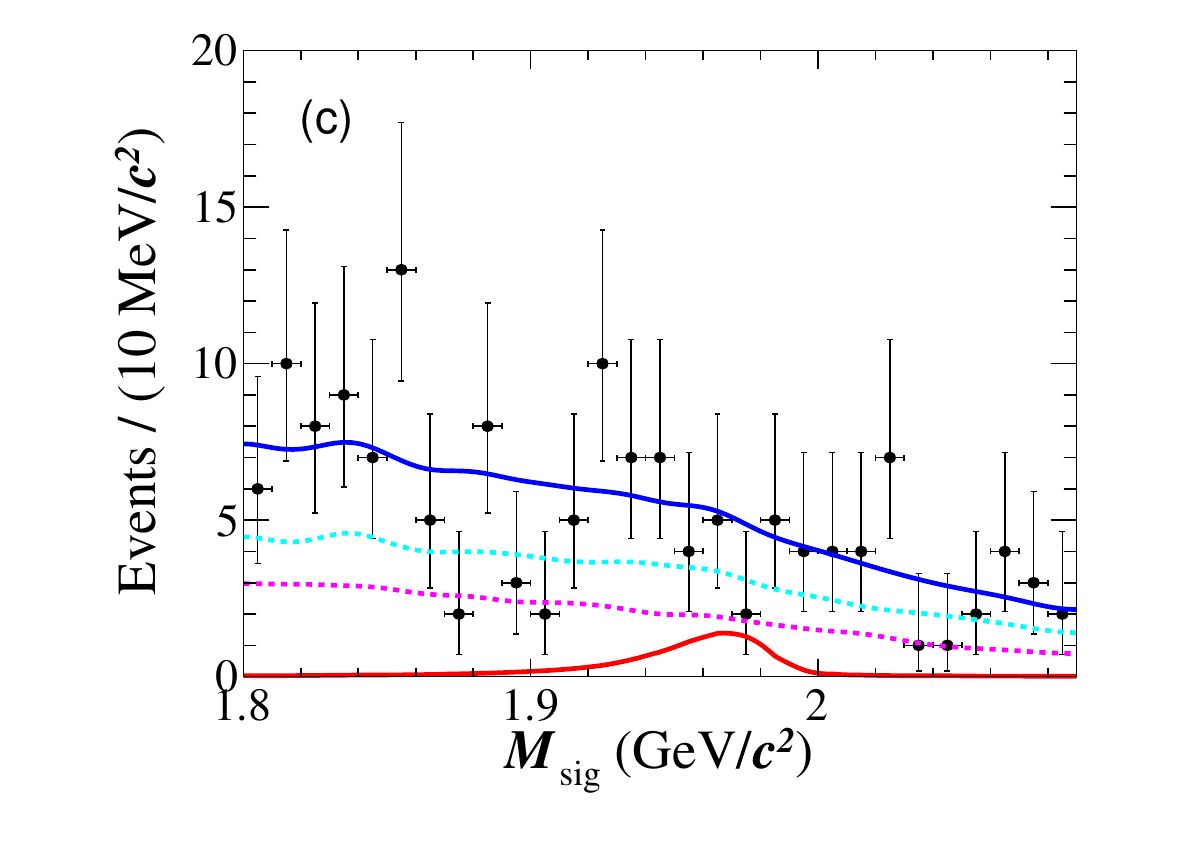}
       \end{minipage}
 \begin{minipage}{7.3cm}      
    \includegraphics[width=7.3cm]{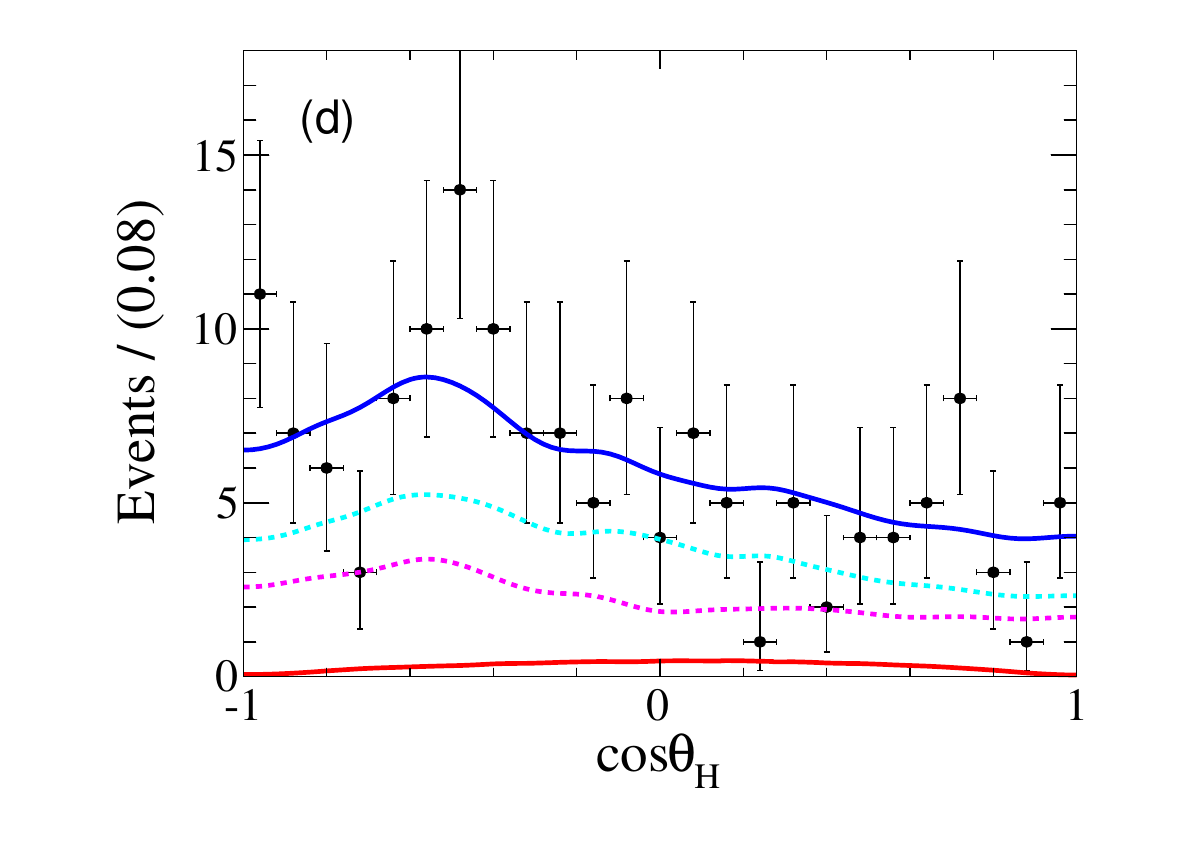}
  \end{minipage} 
		\caption{The distributions of the $M_{\rm sig}$ and cos$\theta_{\rm H}$ for DT candidate events of $D_s^+\to\gamma K^*(892)^+$ via $K^*(892)^+\to K^+\pi^0$~(a)(b) and $K^*(892)^+\to K_S^0\pi^+$~(c)(d) at $\sqrt{s}= 4128-4226$ MeV.
The dots with error bars are data, 
the solid blue lines are the total fit projections, 
the solid red lines represent the signal shape scaled up by a factor of 200, 
the magenta dashed lines represent the ${B}_{D_s^{+}}$ and the cyan dashed lines represent the ${B}_{\rm{other}}$ obtained from the inclusive MC samples.}
\label{2dfit3}
\end{figure}

\begin{table}[tp]
  \centering
  \begin{tabular}{l r@{$\pm$}l r@{$\pm$}l r@{$\pm$}l r@{$\pm$}l}
  \hline
    Tag mode   & \multicolumn{2}{c}{$\epsilon^{\rm DT}_{\rm I} (\%)$} &\multicolumn{2}{c}{$\epsilon^{\rm DT}_{\rm II} (\%)$}    &\multicolumn{2}{c}{$\epsilon^{\rm DT}_{\rm III} (\%)$} &\multicolumn{2}{c}{$\epsilon^{\rm DT}_{\rm IV} (\%)$}\\
  \hline
    $D^-_s\to K^0_SK^-$           &11.94 &0.25 &11.45 &0.11 &11.42 &0.14 &11.03 &0.27\\
    $D^-_s\to K^+K^-\pi^-$        &9.77 &0.10  &9.54 &0.04 
    &9.51 &0.05  &9.30 &0.11\\
	$D^-_s\to K^0_S K^+\pi^-\pi^-$
    &4.82 &0.15  &5.17 &0.07  
    &5.11 &0.09  &5.18 &0.17\\
    \hline
    $D^-_s\to K^0_SK^-$           &10.47 &0.23 &10.20 &0.11 &10.17 &0.13 &10.59 &0.26\\
    $D^-_s\to K^+K^-\pi^-$        &8.05 &0.09  &8.07 &0.04 
    &7.96 &0.05  &7.93 &0.10\\
	$D^-_s\to K^0_S K^+\pi^-\pi^-$
    &3.10 &0.12  &3.17 &0.05  
    &3.13 &0.07  &3.23 &0.13\\
    \hline
  \end{tabular}
	\caption{The DT efficiencies~($\epsilon^{\rm DT}_{\alpha, i}$) of 
    $D_s^+\to\gamma K^*(892)^+$ via $K^*(892)^+\to K^+\pi^0$ and $K^*(892)^+\to K_S^0\pi^+$ for the data samples taken at $E_{\rm cm} = 4.128$ and $4.157$ GeV (I), 4.178 GeV (II), from 4.189 to 4.219 GeV (III), and 4.226 GeV (IV).
    The BFs of the sub-particle~($K^0_S,\pi^0$) decays are not included.}
  \label{abs:dteff}
\end{table}


Since no significant signal is observed, we set a 90\% confidence-level upper limit following the procedure of refs.~\cite{Stenson:2006gwf,BESIII:2021tfk}, incorporating both additive and multiplicative systematic uncertainties.
Additive uncertainties are studied by repeating the fit with alternative background shapes and yields and choosing the most conservative limit. 
Multiplicative uncertainties, which affect the signal efficiency estimation, are folded into the likelihood by convolving with a Gaussian of width $\sigma_\epsilon$. The resulting upper limit is
\begin{equation}
L(\mathcal{B}) \propto \int_0^1 L\left(\mathcal{B} \frac{\epsilon}{\epsilon_0}\right) \exp\left[\frac{-\left(\frac{\epsilon}{\epsilon_0} - 1\right)^2}{2\sigma_\epsilon^2}\right] d\epsilon,
\end{equation}
where \( L(\mathcal{B}) \) is the likelihood distribution as a function of BF; \( \epsilon \) is the expected efficiency and \( \epsilon_0 \) is the averaged MC-estimated efficiency. 
The likelihood curves are shown in Figure~\ref{fig:bf2}.
The averaged efficiency of $D_s^{+}\to\gamma K^*(892)^+$ is $(21.76\pm0.04)\%$.
The upper limit on the BF of $D^+_s\to\gamma K^*(892)^+$ is obtained to be $2.3\times10^{-4}$ at the $90\%$ confidence level.

\begin{figure}[htbp]
          \centering
          \includegraphics[width=10.0cm]{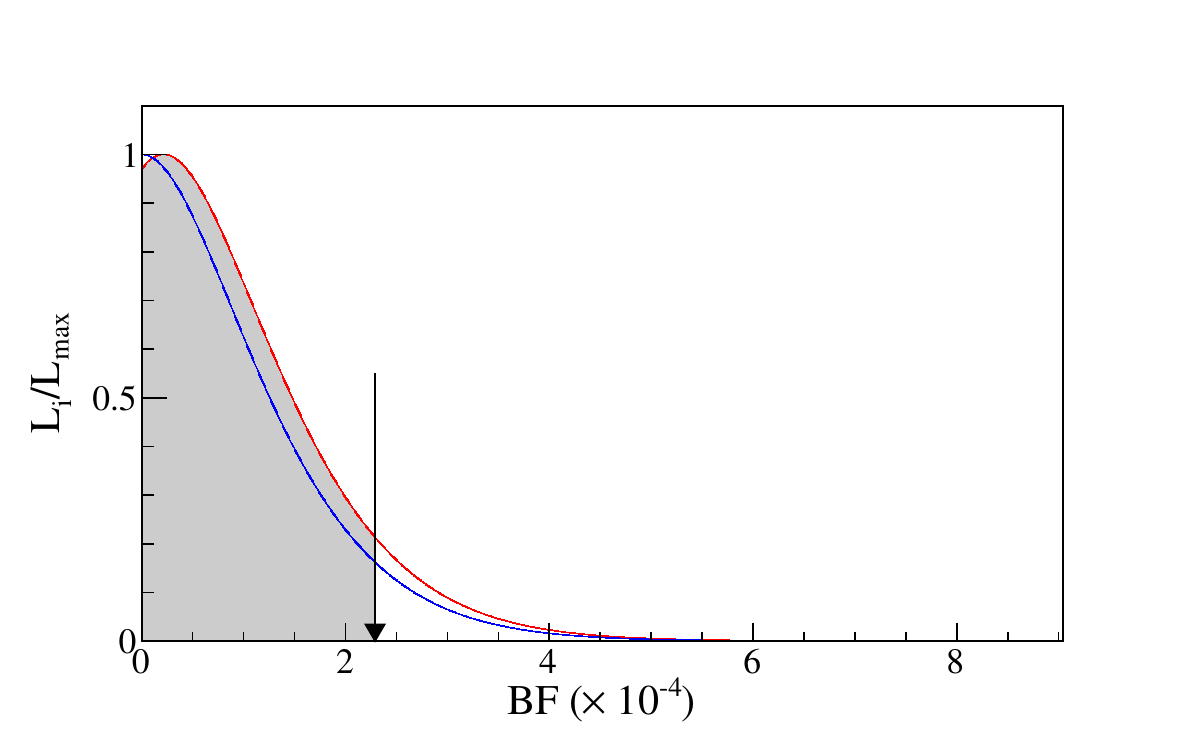}
		\caption{
		Normalized likelihood with respect to the BF of $D^+_s\to\gamma K^*(892)^+$. 
		The blue solid line is the likelihood curve for the nominal fit model, while the red solid line represents the likelihood curve that gives the upper limit after incorporating the systematic uncertainty. 
		The black arrow indicates the result corresponding to the $90\%$ confidence level.
		}
    \label{fig:bf2}
\end{figure}

\section{Systematic uncertainty}
Systematic uncertainties in the BF measurement are categorised as additive and multiplicative ones. 
The additive systematic uncertainties include the signal shape modelling (fits without Gaussian convolution); for the $D_s^{*\pm}D_s^{\mp}$ background, the peaking contribution from $D_s^+ \to \pi^+ \pi^0 \eta$ decays (varied by $\pm 1\sigma$ of its uncertainty), the remaining $D_s^{*\pm}D_s^{\mp}$ background components (varied by $\pm 30\%$); and the non-$D_s^{*\pm}D_s^{\mp}$ background composition (varied by $\pm 30\%$~\cite{BESIII:2021xox}). 
The most conservative limit obtained from these variations is taken.

The multiplicative systematic uncertainties are discussed below.
\begin{itemize}
\item ST yield. 
The total ST yield of all three ST modes is $378,595\pm1,119$,
		resulting in the statistical uncertainty of $\sqrt{1,119^2-378,595}/378,595 = 0.2\%$.
Here, we only consider the statistical fluctuation related to the background of the ST side,
as it is not correlated with the DT sample directly. 

\item Tracking and PID. The $\pi^+$ tracking and PID efficiencies are studied with $e^+e^-\to K^+K^-\pi^+\pi^-$ events. 
The systematic uncertainties are assigned as the data-MC differences of the $\pi^+$ PID and tracking, which are both $1.0\%$~\cite{BESIII:2020ctr}. 

\item $\gamma$ reconstruction. The $\gamma$ reconstruction efficiency
is studied with the control sample of $J/\psi \to \pi^0\pi^+\pi^-$~\cite{BESIII:2011ysp}. The uncertainty due to the $\gamma$ reconstruction is assigned as $1.0\%$.

\item $\pi^0$ reconstruction. The $\pi^0$ reconstruction efficiency is investigated by using a control sample of the process $e^+e^-\to K^+K^-\pi^+\pi^-\pi^0$~\cite{BESIII:2023mie}. 
The uncertainty due to the $\pi^0$ reconstruction is assigned as $2.0\%$.

\item $K_S^0$ reconstruction. The systematic uncertainty due to $K_S^0$ reconstruction is estimated from the measurements of $J/\psi \to K^*(892)^{\mp}K^{\pm}$ and $J/\psi \to \phi K_S^0K^{\pm}\pi^{\mp}$ control samples and found to be 1.5$\%$ per $K_S^0$~\cite{BESIII:2015jmz}.

\item MC statistics. The uncertainty due to the limited signal MC sample size is obtained by $\sqrt{\sum_i{(\frac{f_i\delta_{\epsilon_i}}{\epsilon_i})^2}}$, where $f_i$ is the proportion of each ST yield to the total ST yield, and $\epsilon_i$ and $\delta_{\epsilon_i}$ are the signal efficiency and the corresponding uncertainty of ST mode $i$, respectively. 
This uncertainty is found to be $0.4\%$.

\item $M_{\gamma\gamma_{\rm h}}$ requirement and $M_{\gamma\gamma_{\rm extra}}$ requirement. The systematic uncertainties due to the $M_{\gamma\gamma_{\rm h}}$ requirement and $M_{\gamma\gamma_{\rm extra}}$ requirement are studied using a control sample of $D_s^+\to\pi^+\pi^0 \eta$~\cite{BESIII:2019jjr}. 
The differences in the efficiencies of the $M_{\gamma\gamma_{\rm h}}$ requirement and $M_{\gamma\gamma_{\rm extra}}$ requirement between data and MC simulation, $2.1\%$ and $0.5\%$, are taken as the corresponding systematic uncertainties.

\item {$K^*(892)^+$ mass requirement}.
The systematic uncertainty from the $M_{K^+\pi^0}$ requirement is estimated to be 0.6\%, with the control sample of $\bar{D}^0\to K^+\pi^0 e^-\bar{\nu}_e$~\cite{BESIII:2018qmf}. The systematic uncertainties due to the $M_{K_S^0\pi^+}$ is studied using a control sample of $\bar{D}^0\to K_S^0\pi^+ e^- \bar{\nu}_e$. 
The differences in the efficiencies of the $M_{K_S^0\pi^+}$ between data and MC simulation, $0.2\%$, are taken as the corresponding systematic uncertainties.

\item $E_{\gamma}$ requirement.
The signal loss due to the $E_{\gamma}$ cut is extremely minimal and does not affect the efficiency, so the systematic uncertainty of the $E_{\gamma}$ requirement is neglected.

\item {Contributing BFs}.
The systematic uncertainties of the BFs of $K^*(892)^+\to K^+\pi^0$, $\pi^0\to\gamma\gamma$ and $K^*(892)^+\to K_S^0\pi^+$ are negligible with respect to the other systematic contributions. 
The uncertainty of the BF of $K_S^0\to\pi^+\pi^-$ is 0.1$\%$~\cite{PDG}.
\end{itemize} 

The total multiplicative systematic uncertainty is obtained by adding the individual components in quadrature. Table~\ref{abs:bfsys} summarises the sources of the systematic uncertainties in the BF measurements.

\begin{table}[tp]
  \centering
   \begin{tabular}{l|c|c|c}
  \hline
    Category &Source   &$K^*(892)^+\to K^+\pi^0$ &$K^*(892)^+\to K_S^0\pi^+$\\
  \hline
    \multirow{10}{*}{Multiplicative}&ST yield      &0.2 &0.2\\
    &Tracking                                &1.0 &1.0 \\
    &PID                                     &1.0 &1.0\\
    &$\pi^0$ and $\gamma$ reconstructions     &3.0 &1.0\\
     &$K_S^0$ reconstruction    &- &1.5\\
    
    &MC statistics                           &0.4 &0.4\\
    &$M_{\gamma\gamma_{\rm h}}$ requirement     &2.1 &-\\
    &$M_{\gamma\gamma_{\rm extra}}$ requirement &0.5 &0.5\\
    &$K^*(892)^+$ mass requirement                     &0.6 &0.2\\
    &Contributing BFs &- &0.1\\
    
    \cline{2-4}
    &Total                                   &4.0 
    
    &2.4\\
    \hline
  \end{tabular}
	\caption{Systematic uncertainties~(in unit of $\%$) in the BF measurement of $D^+_s\to\gamma K^*(892)^+$ versus $K^*(892)^+\to K^+\pi^0$ and $K^*(892)^+\to K_S^0\pi^+$.}
  \label{abs:bfsys}
\end{table}

\section{Summary}
Using 7.33~fb$^{-1}$ of $e^+e^-$ collision data collected with the BESIII detector at $E_{\rm cm}=4.128$--4.226~GeV, we have performed the first search for the radiative decay $D_s^+\to\gamma K^*(892)^+$ via $K^*(892)^+\to K^+\pi^0$ and $K^*(892)^+\to K_S^0\pi^+$.  No significant signal is observed.  The upper limit
\begin{equation}
{\cal B}(D_s^+\to\gamma K^*(892)^+)<2.3\times10^{-4}\quad\mbox{(90\% C.L.)}
\end{equation}
is obtained.  Table~\ref{tab:branching_fractions} compares this result with theoretical predictions.  Although the limit is above the predicted range, it approaches the upper end of the theoretical expectations.  None of the models can yet be excluded. The data collected by the Belle Collaboration is expected to be sufficient for this search~\cite{Ipsita:2024wpo}. However, future facilities such as the Super Tau-Charm Facility~\cite{Achasov:2023gey} and Belle II~\cite{Belle-II:2018jsg} will provide larger data samples, enabling more stringent tests. 
 
\begin{table}[tp]
\centering
\begin{tabular}{c|c}
\hline
Theory or experiment & $D_s^+\to\gamma K^*(892)^+$ ($\times 10^{-4}$)  \\
\hline
HSI+WA~\cite{deBoer:2017que}                   & $1.0 - 1.4$ \\
LCSR~\cite{Lyon:2012fk}
    & 0.17  \\
Hybrid~\cite{Fajfer:1998dv,Fajfer:1997bh}  & $0.1 - 0.5$ \\
VMD~\cite{Burdman:1995te}  & $0.1 - 0.3$ \\
\hline
This work & $<2.3$ \\
\hline
\end{tabular}
\caption{The predicted BFs of $D_s^+\to\gamma K^*(892)^+$ from various theories, and the upper limit in this work.}
\label{tab:branching_fractions}
\end{table}

\acknowledgments
The BESIII Collaboration thanks the staff of BEPCII (https://cstr.cn/31109.02.BEPC) and the IHEP computing center for their strong support. This work is supported in part by National Key R\&D Program of China under Contracts Nos. 2023YFA1606000, 2023YFA1606704; National Natural Science Foundation of China (NSFC) under Contracts Nos. 12405094, 12275067, 11635010, 11935015, 11935016, 11935018, 12025502, 12035009, 12035013, 12061131003, 12192260, 12192261, 12192262, 12192263, 12192264, 12192265, 12221005, 12225509, 12235017, 12342502, 12361141819; 
Outstanding Young Talent Support Program of Henan Province;
Science and Technology R$\&$D Program Joint Fund Project of Henan Province  (Grant No.225200810030);
Science and Technology Innovation Leading Talent Support Program of
Henan Province;
the Excellent Youth Foundation of Henan Scientific Commitee under Contract No.~242300421044;
Supported by the China Postdoctoral Science Foundation under Grant Number 2025M783403;
the Chinese Academy of Sciences (CAS) Large-Scale Scientific Facility Program; the Strategic Priority Research Program of Chinese Academy of Sciences under Contract No. XDA0480600; CAS under Contract No. YSBR-101; 100 Talents Program of CAS; The Institute of Nuclear and Particle Physics (INPAC) and Shanghai Key Laboratory for Particle Physics and Cosmology; ERC under Contract No. 758462; German Research Foundation DFG under Contract No. FOR5327; Istituto Nazionale di Fisica Nucleare, Italy; Knut and Alice Wallenberg Foundation under Contracts Nos. 2021.0174, 2021.0299, 2023.0315; Ministry of Development of Turkey under Contract No. DPT2006K-120470; National Research Foundation of Korea under Contract No. NRF-2022R1A2C1092335; National Science and Technology fund of Mongolia; Polish National Science Centre under Contract No. 2024/53/B/ST2/00975; STFC (United Kingdom); Swedish Research Council under Contract No. 2019.04595; U. S. Department of Energy under Contract No. DE-FG02-05ER41374.

\bibliographystyle{JHEP}
\bibliography{references}

\clearpage
\appendix
M.~Ablikim$^{1}$\BESIIIorcid{0000-0002-3935-619X},
M.~N.~Achasov$^{4,c}$\BESIIIorcid{0000-0002-9400-8622},
P.~Adlarson$^{81}$\BESIIIorcid{0000-0001-6280-3851},
X.~C.~Ai$^{87}$\BESIIIorcid{0000-0003-3856-2415},
C.~S.~Akondi$^{31A,31B}$\BESIIIorcid{0000-0001-6303-5217},
R.~Aliberti$^{39}$\BESIIIorcid{0000-0003-3500-4012},
A.~Amoroso$^{80A,80C}$\BESIIIorcid{0000-0002-3095-8610},
Q.~An$^{77,64,\dagger}$,
Y.~H.~An$^{87}$\BESIIIorcid{0009-0008-3419-0849},
Y.~Bai$^{62}$\BESIIIorcid{0000-0001-6593-5665}
% ...
\\
\vspace{0.2cm}
(BESIII Collaboration)\\
\vspace{0.2cm} {\it
$^{1}$ Institute of High Energy Physics, Beijing 100049, People's Republic of China\\
% ...
$^{4}$ Budker Institute of Nuclear Physics SB RAS (BINP), Novosibirsk 630090, Russia\\
% ...
$^{31}$ INFN Sezione di Ferrara, (A)INFN Sezione di Ferrara, I-44122, Ferrara, Italy; (B)University of Ferrara, I-44122, Ferrara, Italy\\
% ...
$^{39}$ Johannes Gutenberg University of Mainz, Johann-Joachim-Becher-Weg 45, D-55099 Mainz, Germany\\
% ...
$^{62}$ Southeast University, Nanjing 211100, People's Republic of China\\
% ...
$^{64}$ State Key Laboratory of Particle Detection and Electronics, Beijing 100049, Hefei 230026, People's Republic of China\\
% ...
$^{77}$ University of Science and Technology of China, Hefei 230026, People's Republic of China\\
% ...
$^{80}$ University of Turin and INFN, (A)University of Turin, I-10125, Turin, Italy; (B)University of Eastern Piedmont, I-15121, Alessandria, Italy; (C)INFN, I-10125, Turin, Italy\\
$^{81}$ Uppsala University, Box 516, SE-75120 Uppsala, Sweden\\
% ...
$^{87}$ Zhengzhou University, Zhengzhou 450001, People's Republic of China\\

\vspace{0.2cm}
$^{\dagger}$ Deceased\\
% ...
$^{c}$ Also at the Novosibirsk State University, Novosibirsk, 630090, Russia\\
% ...

}

\end{document}